\documentclass{article}

\usepackage{arxiv}

\usepackage[utf8]{inputenc} 
\usepackage[T1]{fontenc}    
\usepackage{hyperref}       
\usepackage{url}            
\usepackage{booktabs}       
\usepackage{amsfonts}       
\usepackage{nicefrac}       
\usepackage{microtype}      
\usepackage{amsmath,amssymb}
\usepackage{graphicx}

\title{Hyperspectral phase imaging based on denoising in complex-valued eigensubspace}

\author{
  Igor Shevkunov \\
  Faculty of Information Technology and Communication Sciences\\
  Tampere University\\
  Tampere, Korkealunkatu 10 \\
	and Department of Photonics and Optical Information Technology\\
  ITMO University\\
  St. Petersburg, Russia  \\
  \texttt{igor.shevkunov@tuni.fi} \\
   \And
Vladimir Katkovnik \\
  Faculty of Information Technology and Communication Sciences\\
  Tampere University\\
  Tampere, Korkealunkatu 10 \\
  \texttt{vladimir.katkovnik@tuni.fi} \\
   \AND
  Daniel Claus \\
  Institut f{\"u}r Lasertechnologien in der Medizin und Messtechnik \\
  Helmholtzstraße 12, 89081 Ulm, Germany \\
  \texttt{daniel.claus@ilm-ulm.de} \\
  \And
  Giancarlo Pedrini \\
   Institut f{\"u}r Technische Optik (ITO), Universit{\"a}t Stuttgart \\
  Pfaffenwaldring 9, 70569 Stuttgart, Germany \\
   \texttt{pedrini@ito.uni-stuttgart.de} \\
   \And
Nikolay V. Petrov \\
  Department of Photonics and Optical Information Technology\\
  ITMO University\\
  St. Petersburg, Russia  \\
  \texttt{n.petrov@niuitmo.ru} \\
     \And
Karen Egiazarian \\
  Faculty of Information Technology and Communication Sciences\\
  Tampere University\\
  Tampere, Korkealunkatu 10 \\
  \texttt{Karen.Egiazarian@tuni.fi} \\
}

\begin{document}
\maketitle

\begin{abstract}
A new denoising algorithm for hyperspectral complex domain data has been developed and studied. 
This algorithm is based on the complex domain block-matching 3D filter including the 3D Wiener filtering stage. The developed algorithm is applied and tuned to work in the singular value decomposition (SVD) eigenspace of reduced dimension. 
The accuracy and quantitative advantage of the new algorithm are demonstrated in simulation tests and in processing of the experimental data.
It is shown that the algorithm is effective and provides reliable results even for highly noisy data.

\end{abstract}


\section{Introduction}

Hyperspectral imaging (HSI) enables the collection and processing of data from a large range of the electromagnetic spectrum. Its main applications are in high-quality and contrast imaging  \cite{Tsagkatakis2016}, chemical material identification \cite{Govender2007}, or process detection \cite{Chin2017}.
HSI retrieves information from images obtained across a wide spectral range and hundreds to thousands of spectral channels.
Conventionally, these images are two-dimensional {2D} 
 and stacked together in {3D}
 cubes, where the first two coordinates are spatial  $(x,y)$ and the third one is for the spectral channel, which is usually represented by the wavelength $\lambda$.

Recently, Hyperspectral digital holography (HS{D}H)~\cite{Naik2014, Claus2018} has been developed, which is originated from the work of K. Itoh et al.\cite{Itoh1990} where the interferometric measurement of the three-dimensional Fourier-image of the diffusely illuminated and self-irradiated 
thermal objects was demonstrated for the first time for visible and near infrared spectral ranges, correspondingly.
A typical HS{D}H setup in these frequency ranges is based on a Fourier-transform spectrometer  \cite{BellFourierSp1972} where instead of single-pixel detector a multi-pixel {sensor} (camera) used for the wavefront intensity registration. {Later similar HSDH techniques have been developed for broadband terahertz pulses~\cite{Bespalov2007, IEEE_TTST-Petrov2016} and even for dissipated radiation for unmodified wireless devices~\cite{holl2017holography}.}

As an advantage HS{D}H recovers not only spectrally resolved intensity (amplitude) information, as in the conventional HSI, but also spectrally resolved phase information. 
{F}or HSI {it opens} a new dimension of investigations with the possibility to recover information about height/thickness/relief and refractive indexes of complex-valued object \cite{Kulya2017, Katkovnik2018a,MultiwaveOE2018}. 
In HS{D}H, the {3D data} cubes are complex valued, i.e. each of the {2D} 
 images for each wavelength is complex-valued having {2D} phase and amplitude images.
An important point here is that these cubes are obtained from indirect observations as solutions of the inverse problem. The latter leads to a serious noise amplification in the resulting hyperspectral cubes. Besides,
due to the high spectral resolution, the energy obtained by sensors is separated between many narrow wave bands and limited in each band.

{Narrowing the subject matter to the visible frequency range, it should be noted that radiation sources used in HSDH as a rule a less directed and weak. Additional, during the propagation through the optical setup spectral contents of the radiation could suffer from various losses  associated with inhomogeneous spectral absorption, chromatic aberrations due to the present of optical elements with dispersion of the refractive index.}
 
Due to {all these} limitation{s}, HSI can be very sensitive to additive noise and various kinds of disturbances. 
It is important especially when the hyperspectral sensor is not electronically stable, and the corruption of the electronic charge can affect the spectral value collection easily, especially in low-intensity illuminations that usually occur at the edges of the spectral range used \cite{feng2018}. {Thus, the problem of noise is a significant obstacle to the development of these techniques.}

For noise suppression, the averaging by the sample mean along the wavelength dimension is used routinely \cite{IEEE_TTST-Petrov2016,holl2017holography,Kulya2017,Obara2017,Kalenkov2017}, but it  may result in oversmoothing of the true signal with averaging over a large range of wavelengths. 
Development and application of more sophisticated algorithms for separate wavelength filtering of phase and amplitude appeared as a more efficient tool \cite{Kalenkov2019}.

In this paper, we present a novel advanced algorithm for denoising of complex-domain hyperspectral (HS) data. 
The proposed algorithm estimates the complex-valued signal and noise correlation matrices and then selects a small sized subset of the eigenvectors that best represents the signal subspace in the least squared error sense. The complex-domain filtering is applied for denoising of this small number of eigenimages. The filtered eigenimages are used for reconstruction of the wavefield for all wavelengths. {We will visually and in terms of root-mean-square error show that the algorithm we propose gives a significantly better quality of the reconstructed image in comparison with alternative modern techniques and spectral averaging approach, successfully working even with hyperspectral data characterized by the {extremely small signal-to-noise ratio}. Separately, mention should be made of the results of a comparison of techniques at the task of a phase object reconstruction with highly variable spectral-selective properties.}

The paper organized as follows. Section~\ref{sec:problem} describes a problem formulation and assumptions needed for its solution. 
In Section~\ref{sec:algorithm} the proposed denoising algorithm and its framework are presented.
In section~\ref{sec:simulations} we present simulation experiments for algorithm parameters selection (subsection~\ref{sec:parameters}), for validating filtering results and comparison with alternative filtering techniques on different objects (subsection~\ref{sec:comparison}). 
The phase imaging results for HS data, that was obtained in HS{D}H experiments are discussed in section~\ref{sec:experiment}.
Final conclusions are in section~\ref{sec:conclusion}.

\section{Problem formulation}
\label{sec:problem}
Let $U(x,y,\lambda )\subset\mathbb{C}^{N\times M}$ be a slice of the complex-valued HS cube of the size $N\times M$ on $(x,y)$ provided a fixed wavelength $\lambda $, and $Q_{\Lambda }(x,y)=\{U(x,y,\lambda ),$ $\lambda
\subset \Lambda \}$, $Q_{\Lambda }\subset \mathbb{C}^{N\times M\times L_{\Lambda }}$, be a whole cube composed of a set of the wavelengths $\Lambda $ with the number of individual wavelengths $L_{\Lambda}$. 
Thus, the total size of the cube is $N\times M\times L_{\Lambda }$ pixels. 

The lines of $Q_{\Lambda }(x,y)$ contain $L_{\Lambda }$ spectral observations corresponding to the scene with coordinates $(x,y)$.
Then, the observations of the hyperspectral denoising problem under the additive noise assumption may be written as:
\begin{equation}
Z_{\Lambda }(x,y)=Q_{\Lambda }(x,y)+\varepsilon _{\Lambda }(x,y)\text{, }
\label{eq:NoisyObservations}
\end{equation}%
where $Z_{\Lambda },Q_{\Lambda }$, $\varepsilon _{\Lambda }\subset \mathbb{C}^{N\times M\times L_{\Lambda }}$ represent the recorded noisy HS data, clean HS data and additive noise, respectively.
Accordingly to the notation for the clean image, the noisy cube can be
represented as $Z_{\Lambda }(x,y)=\{Z(x,y,\lambda ),$ $\lambda \in \Lambda \}
$, $Z_{\Lambda }\subset \mathbb{C}^{N\times M\times L_{\Lambda }}$ with the slices $Z(x,y,\lambda )$.

The denoising problem is formulated as reconstruction of unknown $Q_{\Lambda
}(x,y)$ from given $Z_{\Lambda }(x,y)$.
The properties of the clean HS $Q_{\Lambda }(x,y)$ and the noise $\varepsilon _{\Lambda }(x,y)$ are essential for the algorithm development.
 
The following three assumptions are basic hereafter \cite{Zhuang2018}.
\begin{enumerate}
\item \textit{Similarity} of the HS slices $U(x,y,\lambda )$ for close values of $%
\lambda $ follows from the fact that $U(x,y,\lambda )$ are slowly varying
functions of $\lambda $.
It follows that the spectral lines of $Q_{\Lambda }(x,y)$ of the
length $L_{\Lambda }$ live in a $k$-dimensional subspace with $k\ll
L_{\Lambda }$. Therefore, there is a linear transform $E$ reducing the size
of the cube $Q_{\Lambda }(x,y)$ to the cube of the smaller size. Following 
\cite{Zhuang2018}, we herein term the images associated with this $k$%
-dimensional subspace as eigenimages. A smaller size of this subspace
automatically means a potential to improve the HS denoising being produced
in this subspace.
\item \textit{Sparsity} of HS images $U(x,y,\lambda )$ as functions of $(x,y)$ means that there are bases such that $U(x,y,\lambda )$ can be represented with a small number of items of these bases. 
It is one of the natural and fundamental assumptions for design of modern image processing algorithms.
The sparsity for complex-valued images is quite different from the standard formulation of this concept introduced for real-valued signals. 
The complex-valued variables can be defined by any of the two pairs: amplitude/phase and
real/imaginary values, and elements of these pairs are usually correlated \cite{Kalenkov2017,Katkovnik2017}.
\item  The \textit{noise} $\varepsilon _{\Lambda }(x,y)$ is zero mean Gaussian with
unknown correlation matrix $L_{\Lambda }\times L_{\Lambda }$.
\end{enumerate}
Clean image subspace identification is a crucial first step in the developed algorithm. 
The signal and noise correlation matrices are estimated and then used to select the subset of eigenvectors that best represents the signal subspace in the least mean squared error sense. 

\section{Proposed algorithms}
\label{sec:algorithm}
For clearness of the possible filtering techniques that can be applied to the {complex valued} HS data, we present and demonstrate three types of the algorithms: A) Separate denoising of each slice of HS cube, i.e. slice-by-slice denoising;  B) Joint simultaneous denoising of all slices in HS cube; C) Sliding window denoising of HS cube with joint slice denoising in each window.

\subsection{Separate slice denoising}

The algorithms of this group filter the images of the HS cube for
the each wavelength separately with the result which can be shown as 
\begin{equation}
\hat{U}(x,y,\lambda )=CDBM3D\{Z(x,y,\lambda )\},\text{ }\lambda \in \Lambda ,
\label{Alg_1}
\end{equation}%
where $CDBM3D$ is the abbreviation for Complex Domain Block-Matching 3D
filter   and $\hat{U}(x,y,\lambda )$ is an estimate of the
clean unknown wavefront $U(x,y,\lambda )$. 

The complex domain BM3D algorithm is originated in \cite{Katkovnik-2017-CDBM3D}. This concept has been generalized and a wide class of the complex domain BM3D algorithms has been introduced in \cite{Katkovnik2017} as a MATLAB Toolbox. 
Note, that the MATLAB codes for these algorithms are publicly available \cite{Katkovnik2017,Katkovnik-2017-CDBM3D}.

Here, we use CDBM3D as a generic name of these algorithms. Any of these algorithms can be used in  \eqref{Alg_1}.

These algorithms are a generalization for the complex-domain of the popular
real-valued Block-Matching 3D filters \cite{Dabov2007}. Two points define
the potential advantage of CDBM3D in comparison, in particular, with using
BM3D separately for the phase and amplitude as in \cite{Katkovnik2012,Kalenkov2019}. Firstly, CDBM3D processes the phase and amplitude
jointly taking into consideration the correlation quite usual in most
applications, while separate filtering of amplitude and phase
ignores this correlation. Secondly, the basis functions in BM3D are
fixed, while in CDBM3D they are varying data adaptive making estimation
more precise.

Both types of algorithms BM3D and CDBM3D are based on nonlocal similarity of small patches in slice-images always existing in real-life $U_{\Lambda }(x,y)$. The algorithms look for similar patches in slices $Z(x,y,\lambda )$, identify them and process together. This similarity concept allows to use the powerful modern tools of sparse approximation
allowing to design effective denoising algorithms.

\begin{figure}[t]
\center{\includegraphics[width=0.8\linewidth]{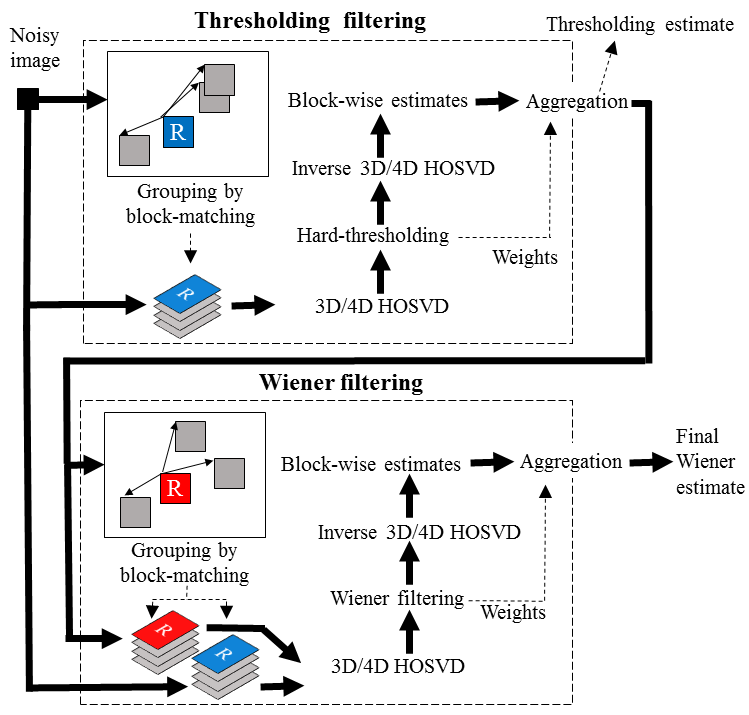}}
\caption{Flow chart of complex domain CDBM3D filters.}
\label{block_scheme}
\end{figure}

The CDBM3D algorithms have a generic structure shown in Fig. \ref{block_scheme} and composed from two successive stages: 'thresholding' and 'Wiener filtering'.
Each of these stages includes: grouping, 3D/4D  High-Order Singular Value Decomposition (HOSVD) analysis, thresholding of HOSVD transforms and aggregation. In the 'Wiener filtering' stage, the thresholding is replaced by Wiener filtering.

Following the procedure in patch-based image processing, the noisy image $%
Z(x,y,\lambda )\subset \mathbb{C}^{N\times M}$ taken with a fixed $\lambda $ is partitioned into small
overlapping rectangular/squares $N_{1}\times M_{1}$ produced for each pixel
of the image. For each patch, we search in $Z(x,y,\lambda )$ for similar
patches, identify them and stack them together in {3D} 
groups (array, tensors). This procedure is called grouping. 

It follows by HOSVD of these groups defining data-adaptive orthonormal
transforms of the complex-valued groups and the core tensors giving the
spectral representation of the grouped data. 

The next step of the algorithm,
as shown in Fig. \ref{block_scheme}, is filtering implemented as
thresholding (zeroing for hard-thresholding) of small items of the core
tensors. Inverse HOSVD using the thresholded core tensors returns block-wise
estimates of the denoised images. These estimates are aggregated in order to
obtain improved image estimates calculated as weighted group-wise estimates.
These grouping, HOSVD, thresholding and aggregation define the so-called thresholding filtering, i.e. the first stage of the algorithm in Fig.\ref{block_scheme}.

The image filtered by thresholding is an input signal of the second stage of
the CDBM3D algorithm - Wiener filtering. The structure of this second part
of the algorithm is similar to the first thresholding part with the only
difference that the thresholding is replaced by the Wiener filtering.

The output of the Wiener filter is the final output of the CDBM3D algorithm. The
details of the threshold and Wiener filtering can be seen in \cite{Katkovnik2017,Katkovnik-2017-CDBM3D,Katkovnik2017b}. It is demonstrated in \cite{Katkovnik-2017-CDBM3D,Katkovnik2017b} that the HOSVD
analysis can be produced using instead of complex-valued variables the pairs
amplitude/phase or real/imaginary values. Then, the groups become {4D} 
and {4D} 
 HOSVD is used for the spectral analysis and filtering.  The flow chart in Fig.~\ref{block_scheme} presents a  structure valid for the both types of
the algorithms where we  show {3D} 
and {4D} 
HOSVD transforms, respectively.

\subsection{Joint slice denoising}

Let us introduce a novel algorithm developed specially for joint processing of slices in HS cubes. We use it with the following
notations:%

\begin{equation}
\hat{U}_{\bar{\Lambda}}(x,y)=\mathcal{CCF}\{Z_{\bar{\Lambda}}(x,y),\text{ }%
\bar{\Lambda}\subset \Lambda \}.  \label{Alg_2}
\end{equation}
Here, $\bar{\Lambda}$ is a set of slices to be denoised.
In particular, for sliding filtering considered later in subsection C, $\bar{\Lambda}$ can be defined as a symmetric wavelength interval centered at $\lambda =\lambda _{0}$ of the width $\delta _{\lambda _{0}}:$%
\begin{equation}
\text{ }\bar{\Lambda}=\{\lambda :\text{ }\lambda _{0}-\delta _{\lambda
_{0}}/2\leq \lambda \leq \lambda _{0}+\delta _{\lambda _{0}}/2\}.
\label{sliding window}
\end{equation}
Complex domain Cube Filter ($\mathcal{CCF}$) processes the data of the cube $%
Z_{\bar{\Lambda}}(x,y)$ jointly and provide the estimates $\hat{U}_{\bar{%
\Lambda}}(x,y)$ for all $\lambda \in $ $\bar{\Lambda}$.

\begin{figure}[t]
\center{\includegraphics[width=0.3\linewidth]{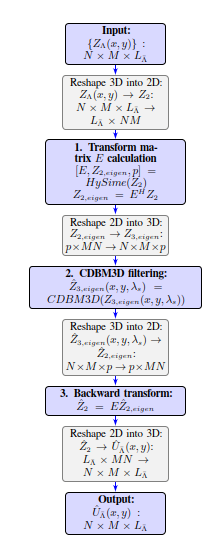}}
\caption{$\mathcal{CCF}$ algorithm. Blue blocks are main steps of the algorithm. Gray blocks are technical steps for data arrays 2D$\leftrightarrows$3D reshaping. }
\label{Fig:block-scheme CCF}
\end{figure}

The $\mathcal{CCF}$ algorithm is presented in Fig.\ref{Fig:block-scheme CCF} and composed from the following steps.

\renewcommand{\labelitemi}{$\circ$}
\begin{itemize}
\item  
Preliminary reshape of {3D} 
 data cube $Z_{\bar{\Lambda}}$\ , $N\times M\times L_{\bar{%
\Lambda}}$\ ,  into the {2D} 
\ matrix $Z$\ of size $L_{\bar{\Lambda}}\times NM:$ $N\times M\times L_{\bar{\Lambda}}\rightarrow L_{\bar{\Lambda}}\times NM$; 
\end{itemize}
\begin{enumerate}
    \item  Calculate the orthonormal transform matrix $E\subset 
\mathbb{C}^{L_{\bar{\Lambda}}\times p}$ and the {2D} 
 transform domain eigenimage\ $Z_{2,eigen}$\  as 
\begin{equation}
\lbrack E,Z_{2,eigen},p]=HySime(Z),  \label{ForwardTransform}
\end{equation}%
where \textit{HySime }stays for Hyperspectral signal Subspace Identification
by Minimum Error \cite{Zhuang2018}, $p$ is a length of the eigenspace.

\textit{HySime} is an important part of the $\mathcal{CCF}$ algorithm. It
identifies an optimal subspace for the HS image representation including
both the dimension of the eigenspace \ $p$ and eigenvectors - columns of $E$. This algorithm is based on the assumption that the slices $U(x,y,\lambda )$ for $\lambda \subset\Lambda $ are realizations of a random field. 

When $E$ is given, the eigenimage is calculated as 
\begin{equation}
Z_{2,eigen}=E^{H}Z.  \label{ForwardTransform1}
\end{equation}%

\end{enumerate}
\begin{itemize}
\item  Reshape\ the 2D transform domain $Z_{2,eigen}$\ , $p\times MN$\ , into
the 3D image domain array $Z_{3,eigen}$ of size $N\times M\times p$ : $%
p\times NM\rightarrow N\times M\times p;$
\end{itemize}

\begin{enumerate}\addtocounter{enumi}{1}
\item  Filter each of the $N\times M$\ 2D images (slices) of $Z_{3,eigen}$ by CDBM3D:
 \begin{equation}
\hat{Z}_{3,eigen}(x,y,\lambda _{s})=CDBM3D(Z_{3,eigen}(x,y,\lambda _{s})).%
\text{ }  \label{filtering}
\end{equation}%
where $p$ eigenvalues $\lambda_s$ belong to the eigenspace.

\end{enumerate}
\begin{itemize}
\item  
Reshape the 3D array $\hat{Z}_{3,eigen}(x,y,\lambda _{s})$ into the 2D
transform domain $\hat{Z}_{2,eigen}$ of size $p\times NM$\ .
\end{itemize}
\begin{enumerate}\addtocounter{enumi}{2}
\item 
Return from the eigenimages of the transform domain to the 2D original image space as follows 
\begin{equation}
\hat{Z}_{2}=E\hat{Z}_{2,eigen}.  \label{BackwardTransform}
\end{equation}

It is the inverse of the transform (\eqref{ForwardTransform1}).

\end{enumerate}
\begin{itemize}
\item  Reshape the 2D image $\hat{Z}_{2}$\ to cube size $N\times M\times L_{%
\bar{\Lambda}}$, it gives the resulting filtered cube $\hat{U}_{\bar{\Lambda}}(x,y)$ (%
\eqref{Alg_2}).
\end{itemize}
These forward and backward reshape passages 2D$\leftrightarrows$3D allow to
define the eigenspace $Z_{2,eigen}$ in the {2D} 
 transform domain
and to produce the CDBM3D filtering in the corresponding {3D} 
 domain $Z_{3,eigen}$ slice-by-slice.
However, in order to return these filtered data $\hat{Z}_{3,eigen}$ into
the original image space we need to use {2D} 
 transform (\eqref{BackwardTransform}) and, thus again to reshape {3D} 
 data into {2D} 
  transform space.

\textit{HySime}, SVD based algorithm, solves the following
problems: estimation of noise and noise covariance matrix and optimization
of the signal subspace minimizing mean squared error between the clean HS $%
U_{\bar{\Lambda}}(x,y)$ and its estimate. Note, that the covariance $%
r_{\lambda ,\lambda ^{\prime }}(x,y)$ averaged over $(x,y)$  and the
the preliminary estimate of HS images (not the final \eqref{BackwardTransform}) is used in this algorithm.
Optimization of the subspace results in minimization of its size and usually the
found subspace dimension $p\ll L_{\bar{\Lambda}}$. It simplifies the
processing of HS data and leads to a faster algorithm. Thus, the CDBM3D
filtering is produced only for $p$ eigenimages but the backward transform (%
\eqref{BackwardTransform}) gives the estimates for all $L_{\bar{\Lambda}}$
spectral images.

The HySime algorithm is a complex domain modification of the HySime
algorithm developed  for real-valued observations in \cite{Bioucas-Dias2008}. More
facts concerning justification of this algorithm as well as motivation
and details can be seen in \cite{Bioucas-Dias2008} because they are nearly identical to those for the complex domain version of the algorithm. 

Overall, the presented $\mathcal{CCF}$ algorithm follows the structure of the fast hyperspectral denoising (FastHyDe) algorithm presented 
for real-valued data in \cite{Zhuang2018}. A generalization to the complex domain required modification of the codes as well as revision of the theoretical background.

In this paper as the CDBM3D algorithm for eigenimage filtering, we use the ImRe-BM3D WI algorithm \cite{Katkovnik2017}. It means that imaginary and real parts of the complex-valued images are used in 4D HOSVD analyses of the 3D grouped data. The algorithm includes both stages: thresholding and Wiener filtering (WI in the algorithm abbreviation). This algorithm enables more efficient performance as compared with the original version of CDBM3D from \cite{Katkovnik-2017-CDBM3D}, where 3D HOSVD is applied to complex-valued images and only thresholding stage is used. 

\subsection{Sliding window CCF}

Statistical modeling and tests of the algorithm show that the
reconstructions $\hat{U}(x,y,\lambda _{s})$ have the accuracy varying with $%
\lambda _{s}$ and the best results are achieved for $\lambda _{s}$ close to
the middle point of the interval $\bar{\Lambda}$.
Sliding $\mathcal{CCF}$ is $\mathcal{CCF}$ applied in the sliding mode with the estimate at the step 3 calculated only for $\lambda _{s}=\lambda _{0}$ in \eqref{sliding window}, where $\lambda _{0}$ takes values from $\Lambda $.
The width of the sliding window $L_{\bar{\Lambda}}$ can be varying with $\lambda _{0}$.

We make publicly available the MATLAB demo-code of the developed $\mathcal{CCF}$ algorithm, see \href{https://tuni-my.sharepoint.com/:u:/g/personal/igor_shevkunov_tuni_fi/EUkd0kMaccRPoyN4yDBfUEcBbXG8VwfZYJGUIls_nQFVyw?e=eGXpt5}{Supplement~1}.  

\section{Simulations}
\label{sec:simulations}
\subsection{Parameters selection}
\label{sec:parameters}
Simulation experiments are produced for the complex-valued HS cube of a transparent phase object. That means that the amplitude A(x,y) of the object wavefront is equal to 1, and the phase is described by the equation:
\begin{equation}
\varphi_{\lambda} (x,y)=2\pi \frac{h(x,y)}{\lambda }(n_{\lambda }-1),
\label{eq:phase}
\end{equation}%
where $\lambda $ is a wavelength of a radiation going through the object,  $n_{\lambda }$ is the refractive index of an object material and $h(x,y)$ is the thickness of the object. 

We model the HS cube for this object by $L_{\Lambda }=200$ slices uniformly covering the wavelength interval $\Lambda=400-798$~nm. 
The refractive index $n_{\lambda }$ is calculated for each $\lambda$ according to  Cauchy's equation with coefficients taken for the  glass BK7 \cite{Glass2014}.
Therefore the noiseless HS cube is represented as 
$Q_{\Lambda }(x,y)=\{U(x,y,\lambda ),$ $\lambda
\subset \Lambda \}$, where $U(x,y,\lambda)$=$A(x,y)\cdot \exp(j\varphi_{\lambda} (x,y))$.

For noise modeling in the HS cube $Z_{\Lambda}(x,y)$ (\eqref{eq:NoisyObservations}), we assume that the additive noise $\varepsilon _{\Lambda }$ is independent and identically distributed complex-valued circular Gaussian with the standard deviation $\sigma$.

We use the  $\mathcal{CCF}$ algorithm in the sliding window mode (Subsection C). In order to select the proper window size $\delta _{\lambda _{0}}$, we produced experiments for the object with the two-peak phase as described in section \ref{sec:interferomic_phase}. The height of these peaks is selected as small in order the phase variations do  not exceed the wrapping range of $[-\pi, \pi)$ for all slices of the HS cube.

The accuracy of phase reconstruction by $\mathcal{CCF}$  is defined as the relative root-mean-squared-error (RRMSE):

\begin{equation}
RRMSE_{\varphi }=\frac{\sqrt{||\hat{\varphi}_{est}-\varphi _{true}||_{2}^{2}}}{\sqrt{||\varphi _{true}||_{2}^{2}}}\text{, } \\
\end{equation}%
where $\hat{\varphi}_{est}$ is a reconstructed phase and  $\varphi_{true}$ is a noiseless phase.

\begin{figure}[t!]
\center{\includegraphics[trim={1cm 0.3cm 0.5cm 0cm},clip,width=1\linewidth]{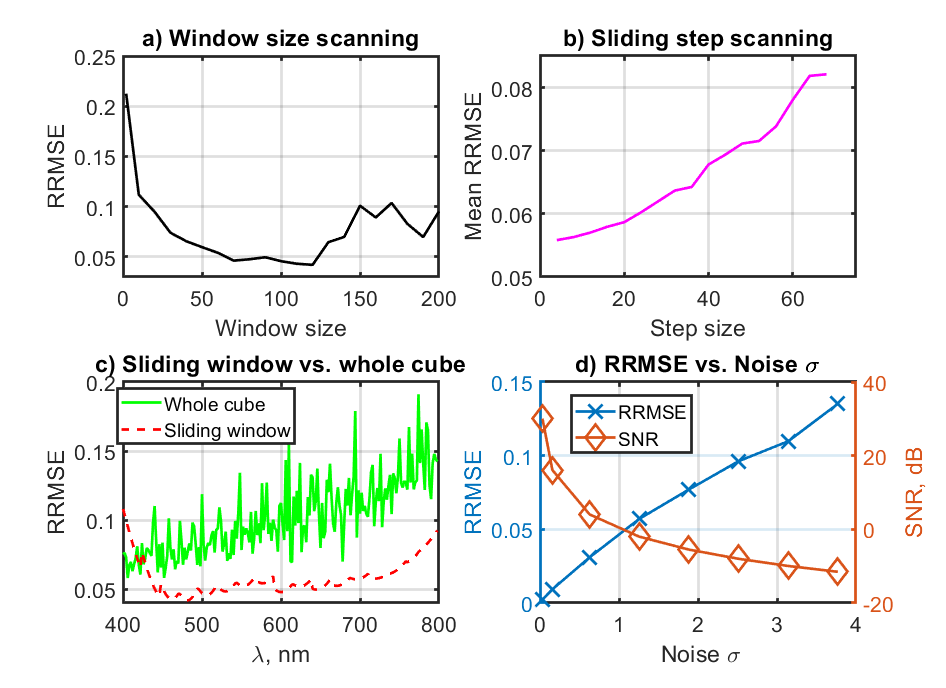}}
\caption{Phase RRMSE curves. (a) RRMSE dependence on window size $L_{\bar{\Lambda}}$; (b) Whole HS cube  mean RRMSE dependence on the sliding step; (c) Comparison of RRMSE distributions for filtering by whole HS cube and by sliding window; (d) RRMSE (blue crosses) and SNR (orange diamonds) curves dependence on noise standard deviation $\sigma$.  }
\label{fig:RRMSEs_for_parameters}
\end{figure}

The RRMSE curves shown in 
 Fig.~\ref{fig:RRMSEs_for_parameters} illustrate our  analysis.   The RRMSE curve in Fig.~\ref{fig:RRMSEs_for_parameters}(a) is obtained for $\lambda_0=598$~nm and calculated for different window size  $L_{\bar{\Lambda}}$.
The best accuracy is achieved for $L_{\bar{\Lambda}}$ somewhere in the interval from 70 to 120.  %
For larger values of $L_{\bar{\Lambda}}$, RRMSE is growing what says that estimation is degrading.
Such RRMSE behavior is due to that initially with growing windows the algorithm takes into account more slices of HS cube, which are similar to the slice of interest $\lambda_0$, therefore it is beneficial for noise suppression and RRMSE curve goes down.
However, if the window is too large  the slices of the HS cube become too different from the slice of interest $\lambda_0$ and are not able to improve the estimation and may even corrupt the resulting filtering.

Let we take $L_{\bar{\Lambda}}=70$ as an optimal window size. According to the idea of the sliding window estimation, the algorithm calculations should be repeated for each slice with this fixed window size. 
However, the neighbouring slices have close values of the wavefronts and the $\mathcal{CCF}$ algorithm with the fixed window size and targeted on $\lambda_0$ can be used in order to calculate the reconstructions also for neighbouring slices. Remind, that this algorithm can give reconstructions for all slices simultaneously. In the sliding window mode of this algorithm we just select a part from this set of the whole estimates.

In this way, we can obtain much faster algorithm for wave field reconstruction using $\mathcal{CCF}$ with the sliding window in step-wise manner, assuming the targeted wavelength $\lambda_0$ is changing step-wise with some step-size and the estimates for intermediate slices are filled by the $\mathcal{CCF}$ reconstruction as reminded above.

Figure~\ref{fig:RRMSEs_for_parameters}(b) shows RRMSE values calculated as the sample mean of RRMSEs for all slices of the HS cube provided the variable step-size for $\lambda_0$.
It is an almost linear dependence, with a bit flatter behavior for small values of the step-size.  
The smallest step-size equal to 1 corresponds to the smallest mean RRMSE value, and as drawback it requires to 200 applications of $\mathcal{CCF}$ to cover all slices of the  HS cube.
As a reasonable trade-off, we took the step-size equal to 12 which corresponds to mean RRMSE smaller than 0.06 but significantly decreases the applications number of $\mathcal{CCF}$ from 200 to 17.

Figure~\ref{fig:RRMSEs_for_parameters}(c) demonstrates a comparison of RRMSE curves for the sliding window step-wise $\mathcal{CCF}$ algorithm with the step-wise equal to 12 (red dotted line) versus  a single run of $\mathcal{CCF}$  (green solid line). 
The advantage of the sliding window technique is obvious with drastically smaller RRMSE values nearly for the whole wavelength interval. Higher values of RRMSE  at the lower and upper bounds of the HS cube can be explained by a non-symmetrical neighborhood for the slice of interest.

Figure~\ref{fig:RRMSEs_for_parameters}(d) illustrates the robustness of the $\mathcal{CCF}$ algorithm regarding the noise level. 
RRMSEs are mainly smaller than 0.1, corresponding to an acceptable quality of filtering. These results are achieved for very low SNR values, down to -8~dB with $\sigma=2.5$.

\begin{figure}[t!]
\center{\includegraphics[trim={0.5cm 2cm 1.0cm 1cm},clip,width=1\linewidth]{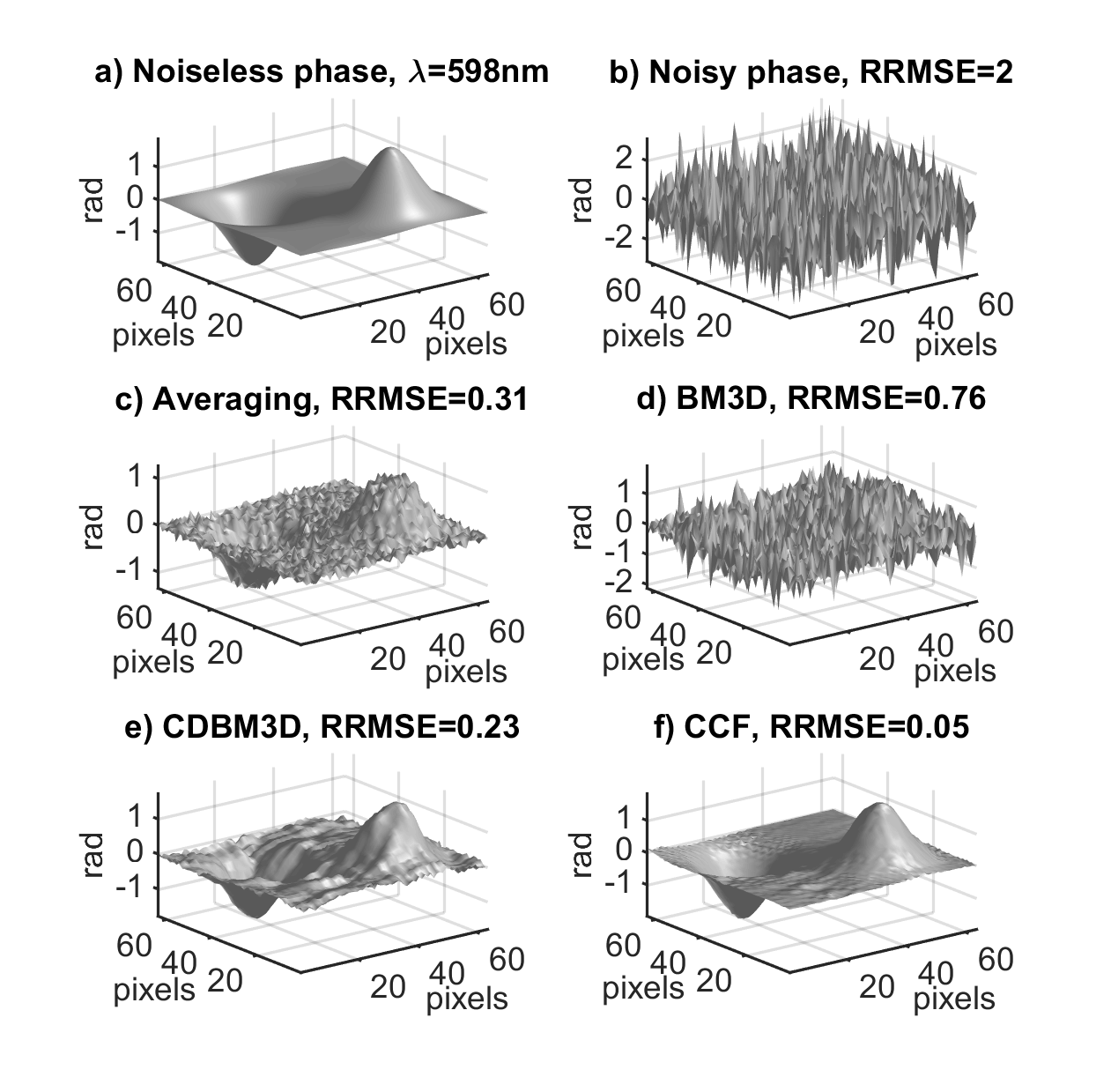}}
\caption{Objects' and filtered phases {not exceeding the range  [$-\pi, \pi$)} for slice $\lambda=598$~nm of  
{the sample}
with corresponding RRMSE values. (a)~Noiseless object phase and (b) with additive noise~$\sigma=1.3$; and phases filtered by: $averaging$ (c), BM3D (d), CDBM3D (e), and $\mathcal{CCF}$ (f). Filtering results for the whole HS cube see in \href{https://tuni-my.sharepoint.com/:v:/g/personal/igor_shevkunov_tuni_fi/ERUo0pWOlT1MgQzKuS65csMBIznKBJn2egZEURblWxpfMQ?e=DZ9FzE}{Supplement~2}.  }
\label{fig:slice_nonwrapped}
\end{figure}

\subsection{Comparison with alternative filtering techniques}
\label{sec:comparison}
To demonstrate the  advantages of the proposed $\mathcal{CCF}$ algorithm we compare filtering results by different state-of-the-art noise suppression methods and $\mathcal{CCF}$ algorithm for different phase objects. These techniques are: $averaging$~\cite{Kalenkov2017}, BM3D~\cite{Dabov2007}, and CDBM3D as in~\cite{Katkovnik2017}. 

$Averaging$ technique calculates the mean value of the thickness $h(x,y)$, which should be the same for each slide of the HS cube, after averaging $h(x,y)$ the phases are calculated back (\eqref{eq:phase}) for comparison.
Since this technique uses the equality of a thickness for the whole cube, it can work only for nonwrapped phases, or otherwise averaging is produced for small number of neighbouring slices. Additionally to account for wavelength-dependent dispersion effect $averaging$ technique needs the knowledge of {dispersion of object's} refractive index 
{~\cite{Kulya2017}}, which is not always possible.

The comparison tests are carried out for objects of the 3 types of phase: ``$interferometric$ simple'', ``$interferometric$ compound'', and ``$wrapped$''.  HS cube for each case is realized by varying the thickness $h(x,y)$ function of the object.
The ''$interferometric$'' assumes that the phase belongs to the interval $[-\pi,\pi)$.
``$Interferometric$ simple''  means that $h(x,y)$ is a smooth function of $(x,y)$ taken such that for whole set of the  wavelengths $\Lambda$ objects' phases do not exceed the range $[-\pi,\pi)$. 

``$Interferometric$ compound'' is a much more complex model. It corresponds to possible non-smooth rapid variations of the thickness $h(x,y)$, which is composed from a few smooth sections with the ``$interferometric$'' phases as in the previous case. However, these sections can be completely different.
In our tests, we assume that there are three this kind of sections. 

This simulation corresponds to objects with different spectral response in $\Lambda$.  
It is an extreme case for spectral analysis, since in real-life such sharp variations of the phase are rare. 
However, it is an interesting test-image to validate the performance of the algorithms.

``$Wrapped$'' means that the phase (absolute phase) corresponding to  $h(x,y)$ may take values beyond the interval $[-\pi,\pi)$ and the algorithm reconstructs the wrapped version of this phase.
It is a difficult object for filtering as phases can be very different from slice to slice.  

For each object, the noisy HS cube is modeled with the noise standard deviation $\sigma$=1.3.

\subsubsection{Interferometric phase simple object}
\label{sec:interferomic_phase}
A first investigated object is modeled with a two-peaks Gaussian phase and an invariant amplitude. The clean phase of the slice corresponding to $\lambda=598~$nm is shown in Fig.~\ref{fig:slice_nonwrapped}(a). 
The corresponding noisy phase of that slice is shown in Fig.~\ref{fig:slice_nonwrapped}(b) and results obtained by by $averaging$, BM3D, CDBM3D, and  $\mathcal{CCF}$  are in Figs.~\ref{fig:slice_nonwrapped}(c-f), respectively.
It is seen from Fig.~\ref{fig:slice_nonwrapped}(b) that the noise level is high, and it is just impossible to trace phase variations behind this noise. 
A significant noise suppression is demonstrated by the $averaging$ algorithm in Fig.~\ref{fig:slice_nonwrapped}(c), where the averaging is produced over all slices of the HS cube. The method works in this test as the phase for the all slices is interferometric and as a result is of a small range of variation in the cube.

\begin{figure}[t!]
\center{\includegraphics[width=0.8\linewidth]{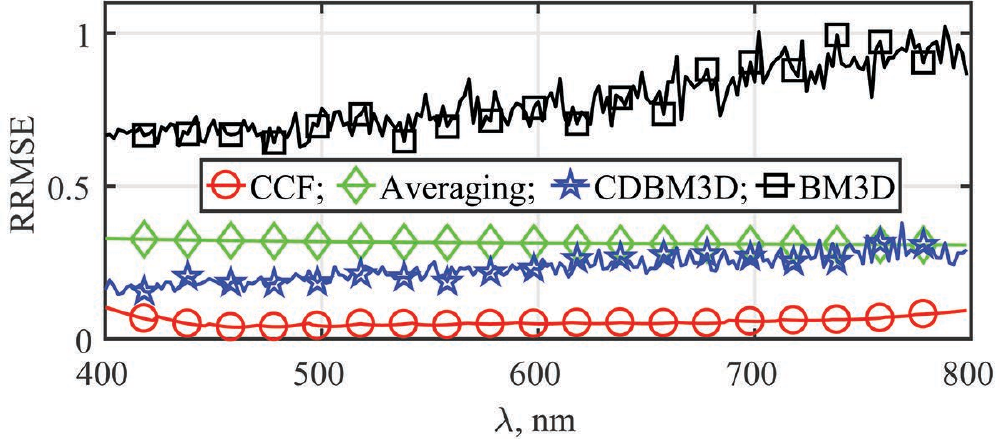}}
\caption{RRMSE distributions for compared filtering algorithms for the object with non-wrapped phase:
red circles curve is for $\mathcal{CCF}$, green diamonds curve is for $averaging$, blue stars is for CDBM3D, black squares curve is for BM3D.}
\label{fig:RRMSEs_All_nonwrapped}
\end{figure}

The BM3D algorithm, Fig.~\ref{fig:slice_nonwrapped}(d),  fails as it is not able to find similar patches for block-matching in the  noisy phase slice shown in Fig.~\ref{fig:slice_nonwrapped}(b). A better filtering is demonstrated by CDBM3D in Fig.~\ref{fig:slice_nonwrapped}(e) due to joint processing of phase and amplitude. Nevertheless, the corresponding RRMSE value is high, the quality of imaging and the accuracy are not acceptable.

In Figure~\ref{fig:slice_nonwrapped}(f), the {result for } $\mathcal{CCF}$ {algorithm} 
is shown{. I}t is seen that noise is suppressed and details of the objects' phase are revealed.   
The video of these filtering results for all cube's slices is available in \href{https://tuni-my.sharepoint.com/:v:/g/personal/igor_shevkunov_tuni_fi/ERUo0pWOlT1MgQzKuS65csMBIznKBJn2egZEURblWxpfMQ?e=DZ9FzE}{Supplement~2}.

Fig.~\ref{fig:RRMSEs_All_nonwrapped} presents RRMSEs for the whole HS cube for each of the compared algorithms. 
The $\mathcal{CCF}$ algorithm demonstrates the best accuracy with the RRMSE values smallest for all slices (wavelengths).  

\begin{figure}[t!]
\center{\includegraphics[trim={2cm 1cm 1.0cm 1cm},clip,width=0.8\linewidth]{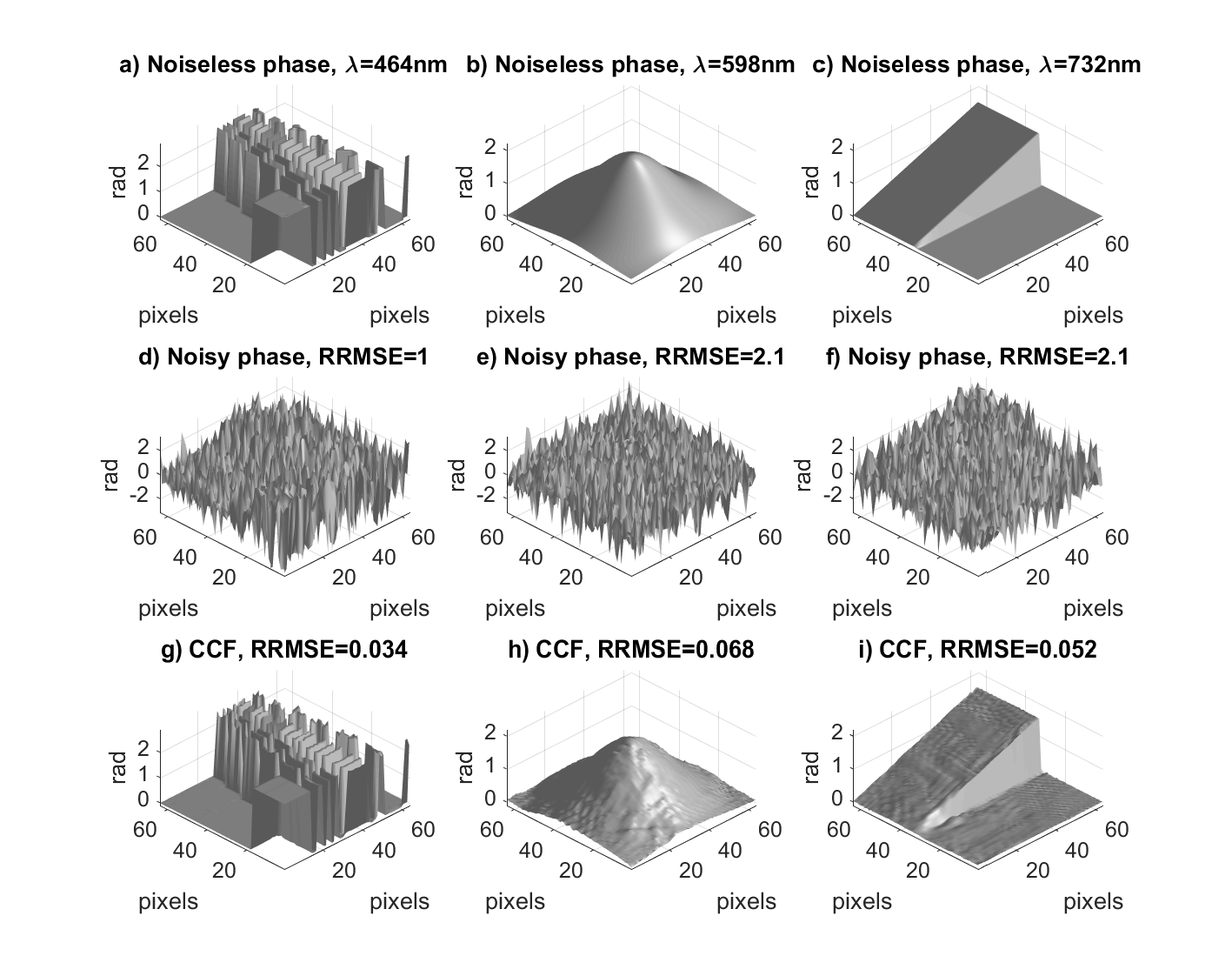}}
\caption{Compound object phases. Noiseless: (a) slice $\lambda=464$~nm is USAF target, (b) slice $\lambda=598$~nm is gaussian peak, (c) $\lambda=732$~nm is inclined surface; images (d,e,f) are corresponding noisy phases with noise standard deviation $\sigma=1.3$; and (g,h,i) are filtered phases by $\mathcal{CCF}$ of corresponding noisy phases. Filtering results for the whole HS cube see in \href{https://tuni-my.sharepoint.com/:v:/g/personal/igor_shevkunov_tuni_fi/EUFYiE6UizJJtT8F6vrw0cUBJROAuOmmSCYq7pngtXz_Xg?e=5CfJGD}{Supplement~3}. }
\label{fig:slice_3object}
\end{figure}

\begin{figure}[b!]
\center{\includegraphics[width=0.8\linewidth]{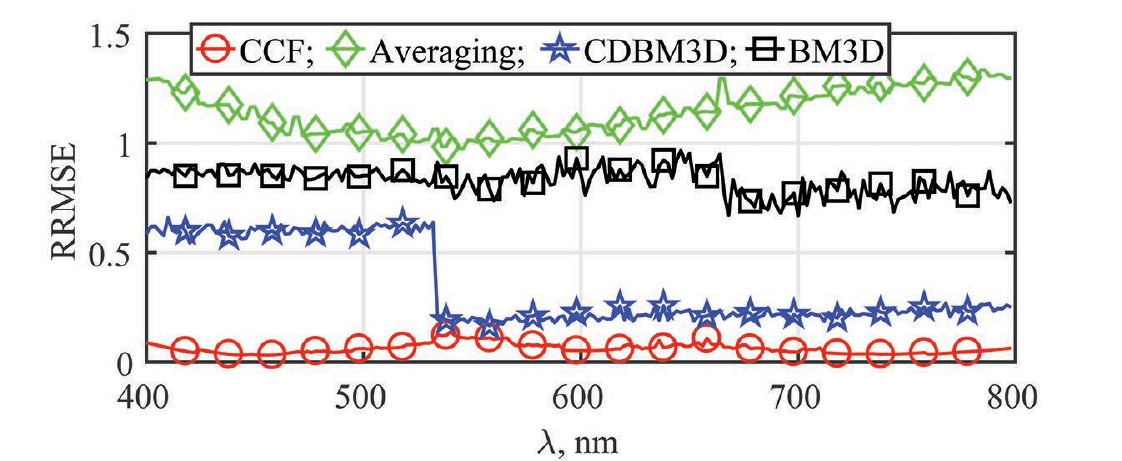}}
\caption{RRMSE distributions for compared filtering algorithms for non-wrapped compound phase object:
red circles curve is for $\mathcal{CCF}$, green diamonds curve is for $averaging$, blue stars is for CDBM3D, black squares curve is for BM3D.}
\label{fig:RRMSEs_All_3objects}
\end{figure}

\subsubsection{Interferometric phase compound object}

The second considered phase object  {wa}s composed from three parts with thicknesses $h(x,y,\lambda_s)$ defined as follows: for $\{\lambda_s, s=[1:\frac{1}{3} L_\Lambda)\}$, $h(x,y)$ is a binary USAF test target; for $\{\lambda_s, s=[\frac{1}{3} L_\Lambda:\frac{2}{3} L_\Lambda)\}$ it is a Gaussian peak, and for $\{\lambda_s, s=[\frac{2}{3} L_\Lambda:L_\Lambda]\}$ it is an inclined discontinuous surface. 
The corresponding three parts of the phase slices are shown in Fig.~\ref{fig:slice_3object}(a-c) as noiseless clean and with the high-level additive noise in Fig.~\ref{fig:slice_3object}(d-f). 

After filtering by the $\mathcal{CCF}$ algorithm working in the sliding window mode, we are obtain results shown in Fig.~\ref{fig:slice_3object}(g-i) for each part of the compound phase corresponding to the noisy images in Fig.~\ref{fig:slice_3object}(d-f). 

Comparing the filtered phases with the clean images in Fig.~\ref{fig:slice_3object}(a-c) we may conclude that the algorithm enables a high quality reconstruction working with compound data successfully separating three completely different phase distributions.
The demo-video showing results for all wavelength, i.e.  for the whole HS cube, is available in \href{https://tuni-my.sharepoint.com/:v:/g/personal/igor_shevkunov_tuni_fi/EUFYiE6UizJJtT8F6vrw0cUBJROAuOmmSCYq7pngtXz_Xg?e=5CfJGD}{Supplement~3}.

The RRMSE curves as functions of wavelength are shown in Fig.~\ref{fig:RRMSEs_All_3objects} for the four compared algorithms. 
The $averaging$ algorithm (green diamonds curve) fails completely because the phase images are very different for different part of the phase object.
CDBM3D and BM3D show different accuracy for different parts of the compound data. 
In particular, CDBM3D  shows a better performance for the parts with plain surfaces: Gaussian peak and inclined surface shaped phase. 
The  $\mathcal{CCF}$ curve (red circles) demonstrates the best performance with smallest RRMSE value more or less invariant with respect to wavelength. Thus, it enables the uniform accuracy for all wavelength despite a great difference between the phase images for different parts of the wavelength interval.
\begin{figure}[t!]
\center{\includegraphics[trim={0.5cm 3cm 0.0cm 1cm},clip,width=0.9\linewidth]{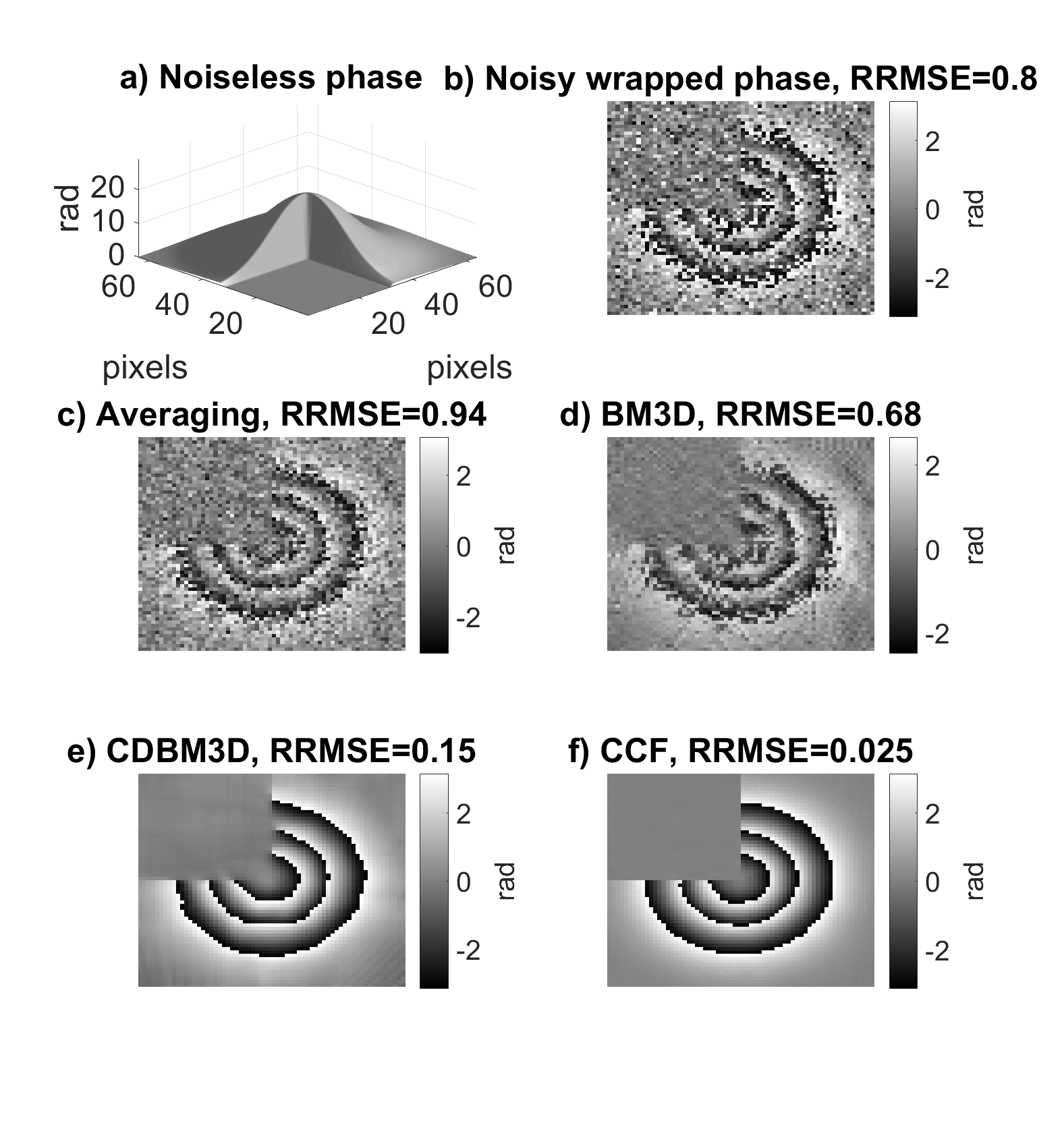}}
\caption{Objects' and filtered phases for slice $\lambda=598$~nm of wrapped object with corresponding RRMSE values. Noiseless absolute (a) and noisy ($\sigma=1.3$) wrapped (b) object phases; and noisy wrapped phases filtered by: $averaging$ (c), BM3D (d), CDBM3D (e), and $\mathcal{CCF}$ (f). Filtering results for the whole HS cube see in \href{https://tuni-my.sharepoint.com/:v:/g/personal/igor_shevkunov_tuni_fi/EdR18qkpiflPm0zFswhmTE0BCA6CEZmH75a-BYoUmX8TLg?e=OKvO9x}{Supplement~4}.}
\label{fig:slice_wrapped}
\end{figure}

\begin{figure}[t]
\center{\includegraphics[width=0.8\linewidth]{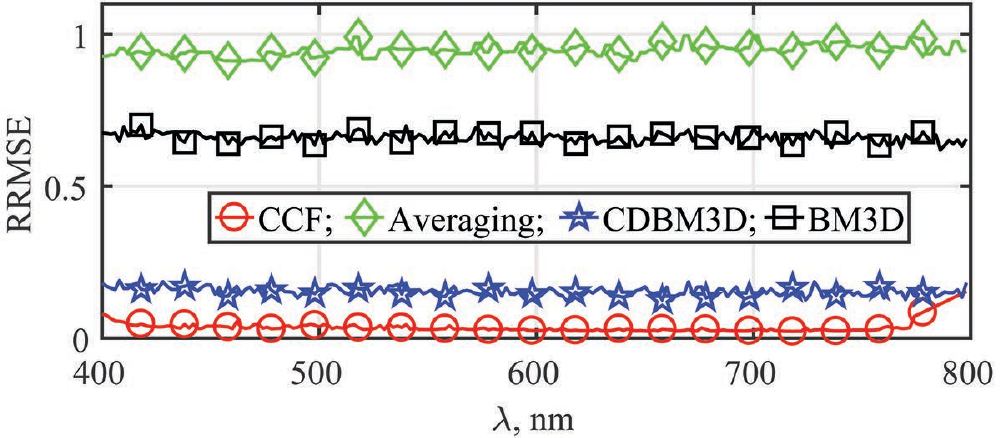}}
\caption{RRMSE distributions for compared filtering algorithms for wrapped phase object:
red circles curve is for $\mathcal{CCF}$, green diamonds curve is for averaging, blue stars is for CDBM3D, black squares curve is for BM3D. }
\label{fig:RRMSE_All_wrappwed}
\end{figure}
\subsubsection{Wrapped phase object}

Consider the phase object provided the phase defined by the truncated Gaussian peak with a maximum phase delay $\varphi_{400nm}=28.9$~rad 
achieved for $\lambda=400$~nm. 
Fig.~\ref{fig:slice_wrapped}(a) shows the shape of the absolute phase corresponding to the middle slice of the HS cube with $\lambda=598$~nm. 

As the phase takes values out of the interval $[-\pi, \pi)$ the observations are defined by not absolute but wrapped phases. We use the tested algorithms for reconstruction of these wrapped phases, which are essentially varying on the wavelength range.

The wrapped noisy phase with $\sigma=1.3$ is presented in Fig.~\ref{fig:slice_wrapped}(b). The filtering results given by the $averaging$, BM3D, CDBM3D, and $\mathcal{CCF}$ algorithms are shown in Figs.~\ref{fig:slice_wrapped}(c-f), respectively.
The $averaging$ algorithm is implemented using a sliding window estimator averaging with the window size equal to 2, i.e. averaging is produced for the pairs of two neighboring slides. The noisy image in Fig.~\ref{fig:slice_wrapped}(c) and high value of RRMSE indicates that the algorithm fails.

The BM3D algorithm is also demonstrate a poor visual performance and poor accuracy. 
CDBM3D produces a much better visual wrapped phase reconstruction but with quite high RRMSE value, thus the accuracy of the wrapped phase reconstruction is not good, Fig.~\ref{fig:slice_wrapped}(e).
$\mathcal{CCF}$ algorithm demonstrate the best performance visually and numerically, 
Fig.~\ref{fig:slice_wrapped}(f).
RRMSE curves for the whole HS cube with comparison of the algorithms see Fig.\ref{fig:RRMSE_All_wrappwed} prove the great advantage 
of the $\mathcal{CCF}$ algorithm.

Demo-video showing the results for all algorithms as well as variation of the object phase over the whole wavelength range can be seen 
\href{https://tuni-my.sharepoint.com/:v:/g/personal/igor_shevkunov_tuni_fi/EdR18qkpiflPm0zFswhmTE0BCA6CEZmH75a-BYoUmX8TLg?e=OKvO9x}{Supplement~4}.

\section{Experimental results}
\label{sec:experiment}

\begin{figure}[t]
\begin{minipage}[]{0.38\linewidth}
\center{\includegraphics[width=0.7\linewidth, height=0.7\linewidth]{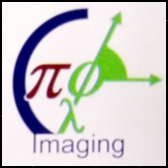} \\ a)}  
\end{minipage}
\hfill
\begin{minipage}[]{0.5\linewidth}
\center{\includegraphics[width=0.7\linewidth]{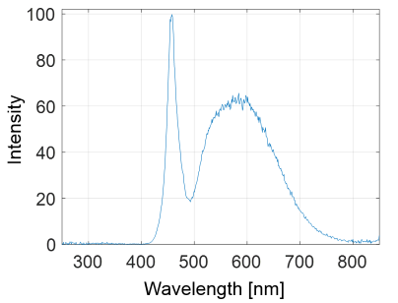} \\ b)}
\end{minipage}
\caption{Transparent object slide 5~mm$\times5$~mm (a) and used LED spectra (b).}
\label{fig:realObj_and_LED}
\end{figure}

The experimental HS data are obtained via spectrally resolved digital holography, as described in \cite{Claus2018}. 
The object is a transparent color slide (Fig.~\ref{fig:realObj_and_LED}(a)) and the light source is a white LED the spectrum of which is shown in Fig.~\ref{fig:realObj_and_LED}(b). 
In spectral regions of low light intensity the SNR of the holograms is low and the corresponding reconstructed wavefronts would be more noisy, and in regions where LED intensity is higher the SNR is higher also, and therefore reconstructed wavefronts could be less noisy.  

\begin{figure}[h]
\center{\includegraphics[width=0.9\linewidth]{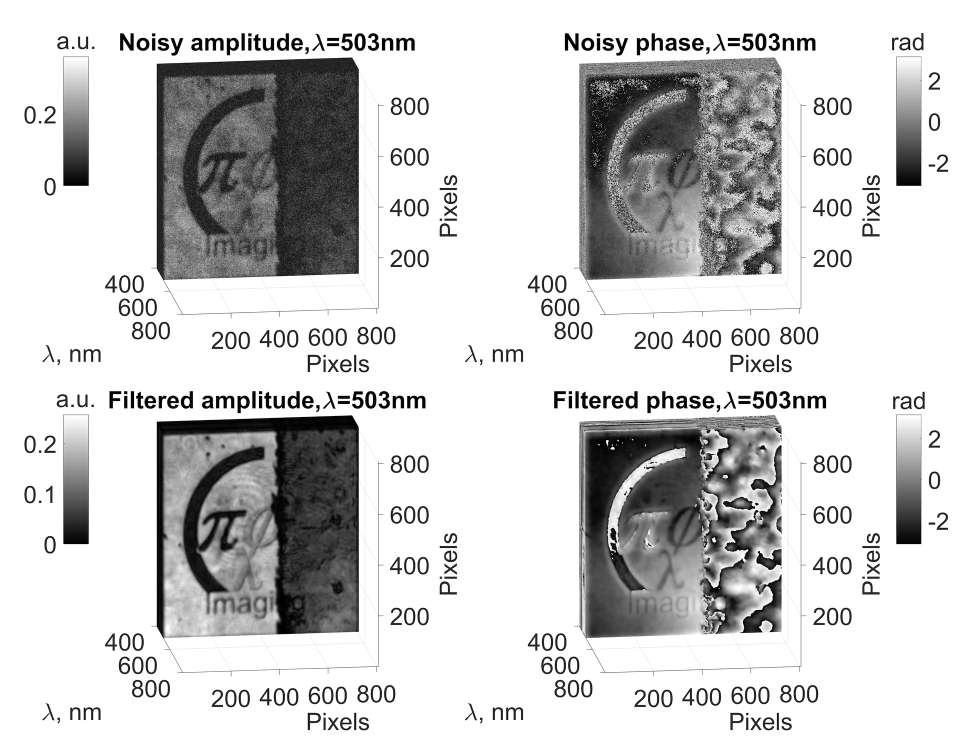}}
\caption{Images of objects' amplitude and phase  corresponding to $\lambda=503$~nm slice of the HS cube. Top row: noisy amplitude (left) and phase (right); bottom row: $\mathcal{CCF}$ filtered amplitude and phase, correspondingly.  Filtering results for the whole HS cube see in \href{https://tuni-my.sharepoint.com/:v:/g/personal/igor_shevkunov_tuni_fi/ERkAPfIy4DpBmntq5tAdiw8BzpnvKxM2eU4WaMHE3OxyKg?e=AoWkZV}{Supplement~5}. }
\label{fig:slice_realobject_503nm}
\end{figure}

\begin{figure}[h]
\center{\includegraphics[width=0.9\linewidth]{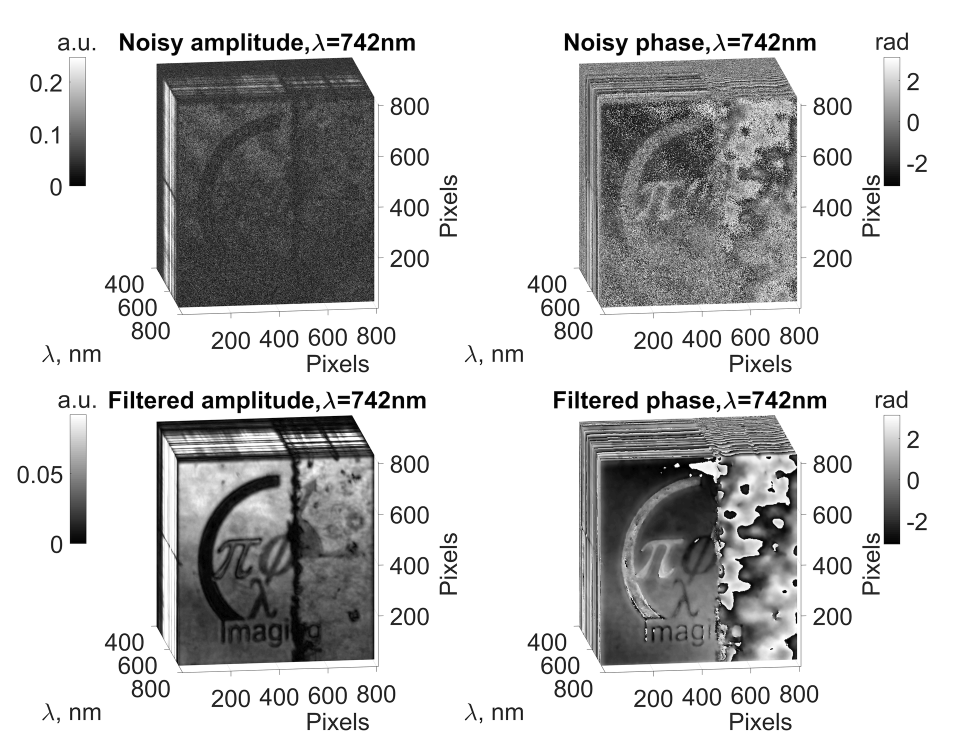}}
\caption{Images of objects' amplitude and phase corresponding to $\lambda=742$~nm slice of the HS cube. Top row: noisy amplitude (left) and phase (right); bottom row: $\mathcal{CCF}$ filtered amplitude and phase, correspondingly.
Filtering results for the whole HS cube see in \href{https://tuni-my.sharepoint.com/:v:/g/personal/igor_shevkunov_tuni_fi/ERkAPfIy4DpBmntq5tAdiw8BzpnvKxM2eU4WaMHE3OxyKg?e=EbbS1I}{Supplement~5}.}
\label{fig:slice_realobject_742nm}
\end{figure}

Examples of amplitudes and phases slices of the observed HS data are shown in the top rows of Figs.~\ref{fig:slice_realobject_503nm} and \ref{fig:slice_realobject_742nm} for amplitudes and phases corresponding to 503~nm (less noisy)  and 743~nm (more noisy), respectively. 
For the less noisy 503~nm slice (top row Fig.~\ref{fig:slice_realobject_503nm}), the structure of the object is clearly seen in both phase and amplitude, while for the noisier slice 743~nm (top row Fig.~\ref{fig:slice_realobject_742nm}) it is hard to distinguish details of the object behind the noise. 
However, after the $\mathcal{CCF}$ filtering the noise is sufficiently suppressed in the whole HS cube.
Object details are revealed for every slice regardless of the noise level, see bottom rows in Figs.\ref{fig:slice_realobject_503nm} and \ref{fig:slice_realobject_742nm}. 
Video-demo with the filtering results for the whole HS cube are presented in \href{https://tuni-my.sharepoint.com/:v:/g/personal/igor_shevkunov_tuni_fi/ERkAPfIy4DpBmntq5tAdiw8BzpnvKxM2eU4WaMHE3OxyKg?e=AoWkZV}{Supplement~5}.

\begin{figure}[h]
\center{\includegraphics[width=1\linewidth]{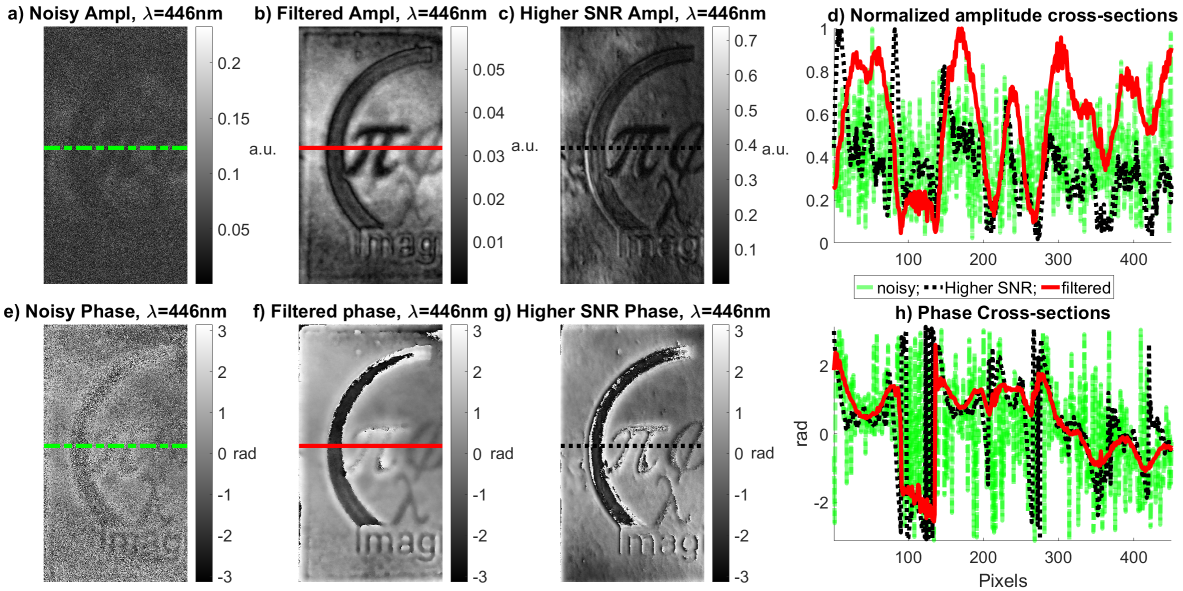}}
\caption{Comparison of noisy amplitude~(a) and phase~(e) of slice corresponding to $\lambda=446$~nm with filtered amplitude~(b) and phase~(f) by $\mathcal{CCF}$ algorithm and with amplitude~(c) and phase~(g) from HS cube with higher SNR. Longitudinal cross-sections for phases and amplitudes are in (d) and (h), respectively. Green curve is for noisy slice, red is for slice filtered by $\mathcal{CCF}$ algorithm, and black is for slice with higher SNR.  }
\label{fig:comparison_real_objects}
\end{figure}

Furthermore, the $\mathcal{CCF}$ filter enables the recovery of information that was nearly completely lost in the observed noisy slices. For a demonstration of the correctness of the $\mathcal{CCF}$ filtering, we compare three slices: 1) the slice 446~nm from noisy non-filtered HS cube with 2) the same noisy slice from the HS cube but filtered by $\mathcal{CCF}$ and 3) the non-filtered 446~nm slice obtained for the same test-object in another experiment with higher SNR.
In Figure~\ref{fig:comparison_real_objects}, the top row is for amplitudes and the bottom row is for phases;  from left to right: the first column is for images of noisy slice, the second column is for $\mathcal{CCF}$ filtered, the third one is for the noisy slice of the higher SNR. In the fourth column, we show the cross-sections of the images shown in the columns 1-3. 
We can see, that the information almost completely lost in the noisy slice (Fig.~\ref{fig:comparison_real_objects}(a,e)) is revealed by the  $\mathcal{CCF}$ algorithm (Fig.~\ref{fig:comparison_real_objects}(b,f)) and these found amplitude/phase features coincide with those clearly seen in the slice of higher SNR (Fig.~\ref{fig:comparison_real_objects}(c,g)). 

The amplitude cross-sections that are presented in Fig.~\ref{fig:comparison_real_objects}(d) are normalized/scaled to the interval $[0,1]$,
since the HS cubes, in the compared two experiments, were obtained with different light intensities.
Nevertheless, the comparison qualitatively confirms that there are no visual artifacts generated by the algorithm and the obtained filtered amplitude/phase changes correspond to more precise observations obtained from the higher SNR experiment.
From the comparison of amplitude cross-sections it can be concluded that for the amplitude filtered by $\mathcal{CCF}$ (red solid curve) all object features are revealed in the same locations as in the amplitude of  the HS cube with higher SNR (black dotted curve).
Different values of the amplitudes in the cross-sections can be referred to different illuminations in the considered two experiments.

The phase cross-sections in Fig.~\ref{fig:comparison_real_objects}(h) also confirm that there are no visual artifacts produced by the algorithm $\mathcal{CCF}$ filtering (red solid curve) and the phase curves are quite close to those obtained in the experiment with higher SNR. 
The noise level in the $\mathcal{CCF}$ curve is even lower than that in the experiment with higher SNR (black dotted curve). 
In both amplitude and phase cross-sections the green curves correspond to the noisy observed slice and indicate very noisy data.

\section{Conclusion}
\label{sec:conclusion}
In this paper, we present the denoising algorithm for hyperspectral complex-valued data. 
Based on comprehensive investigations, we have demonstrated the state-of-the-art performance of the algorithm owing to the SVD analysis of noisy hyperspectral observations and complex domain BM3D filtering in the reduced dimension SVD eigenspace.
The study includes multiple simulation tests and processing of those and HS digital holography data.
The algorithm is robust with respect to noisy data, and produces reliable results even for extra low SNR, down to -8 dB. {It }
demonstrates a stable effective performance for different types of HS data with interferometric and wrapped phases, the latter without involving unwrapping procedures.

We believe that the extremely high performance of our technique of processing interferometric data cubes demonstrated in this work will  eliminate the key constraint preventing the widespread usage of phase HSDH imaging at practice.  In particular, the following areas are attractive for this technique: (i) spatial-spectral analysis of biological products in the visible ultraviolet spectral ranges (as an example, see the work \cite{Soltani2019}); (ii) HS analysis of dispersion and absorption properties in near-infrared Fourier spectroscopy (see details e.g. in \cite{Manley2014}) extended with the addition of spatial degrees of freedom; (iii) Development of new approaches for non-invasive taxonometry of marine plankton based on broadband phase imaging. For such techniques background, see \cite{Grcs2018}; (iv) Analysis of the broadband wavefront propagation dynamics  through the scattering and dispersive media~\cite{Mounaix2016}.  

\section*{Funding}
Academy of Finland, project no. 287150, 2015-2019;
Russian Foundation for Basic Research (18-32-20215/18). 

 


\bigskip \noindent See  \href{https://tuni-my.sharepoint.com/:u:/g/personal/igor_shevkunov_tuni_fi/EUkd0kMaccRPoyN4yDBfUEcBbXG8VwfZYJGUIls_nQFVyw?e=eGXpt5}{Supplement~1},
\href{https://tuni-my.sharepoint.com/:v:/g/personal/igor_shevkunov_tuni_fi/ERUo0pWOlT1MgQzKuS65csMBIznKBJn2egZEURblWxpfMQ?e=DZ9FzE}{Supplement~2},
\href{https://tuni-my.sharepoint.com/:v:/g/personal/igor_shevkunov_tuni_fi/EUFYiE6UizJJtT8F6vrw0cUBJROAuOmmSCYq7pngtXz_Xg?e=5CfJGD}{Supplement~3},
\href{https://tuni-my.sharepoint.com/:v:/g/personal/igor_shevkunov_tuni_fi/EdR18qkpiflPm0zFswhmTE0BCA6CEZmH75a-BYoUmX8TLg?e=OKvO9x}{Supplement~4},
\href{https://tuni-my.sharepoint.com/:v:/g/personal/igor_shevkunov_tuni_fi/ERkAPfIy4DpBmntq5tAdiw8BzpnvKxM2eU4WaMHE3OxyKg?e=AoWkZV}{Supplement~5} for supporting content.


\end{document}